\documentclass[12pt]{article}
\usepackage{epsfig}
\usepackage{cite}

\begin{document}

\title{5,000,000 Delays---Some Statistics\footnote{In: IVS 2004 General Meeting Proc., Eds. N. R. Vandenberg, K. D. Baver, NASA/CP-2004-212255, 2004, 47--51.}}
\author{Zinovy Malkin\footnote{Current affiliation: Pulkovo Observatory, St.~Petersburg, Russia} \\ Institute of Applied Astronomy, St.~Petersburg, Russia}
\date{\vspace{-10mm}}
\maketitle

\begin{abstract}
5,000,000 VLBI delays are stored now in the IVS data base
and available for scientific analysis.
This is a remarkable result of more than 20 years of geodetic
VLBI history.  This paper presents some statistics related
to the VLBI observations during almost 25 years of geodetic
VLBI.
\end{abstract}

At the end of 2004, the number of VLBI delays obtained in the framework
of the geodetic and astrometric VLBI programs reached 5,000,000!
This is one of the major milestones in the VLBI history
resulting from almost 25 years of heroic
efforts by astronomers and physicists, engineers and programmers, network
stations and correlator teams.

Most of the statistics presented here
(when not indicated explicitly) are related to
all the observing sessions, independent of their duration.
As can be seen from Table~\ref{tab:stat_duration}, only number of
sessions substantially depends on the session set taken into account,
other statistics are
practically the same for all sessions and 24h ones (we consider the
session as 24h one if its duration is 18 hours or greater).
One can see that most of observations were obtained during the 24h and
intensives ($<2h$) sessions.  The sessions with duration 2--18h were,
evidently, rather sporadic.

\begin{table}[ht!]
\centering
\caption{Statistics depending on the session duration.
Number of observations is given in thousands.
Number of stations includes also the experimental ones
(KASL, MOJAVLBA, NOTOX, LEFT85\_1, VLBA85\_3, WIDE85\_3).}
\label{tab:stat_duration}
\small
\begin{tabular}{|l|c|c|c|c|c|c|}
\hline
Session duration       & All  & $\ge 18$h & $<18$h & $<12$h & $<6$h & $<2$h \\
\hline
Number of observations & 5005 & 4913      & 92     & 80     & 68    & 64    \\
Number of sessions     & 8528 & 3757      & 4771   & 4737   & 4633  & 4528  \\
Number of stations     & 159  & 156       & 53     & 40     & 29    & 17    \\
Number of baselines    & 1356 & 1335      & 167    & 94     & 51    & 33    \\
Number of sources      & 2254 & 2248      & 406    & 355    & 201   & 119   \\
\hline
\end{tabular}
\end{table}

These observations were collected in 8528 sessions (3757 of them with duration
18h or greater) at 159 stations including experimental ones, on 1356
baselines (there was a misprint in the IVS Newsletter of August 2002,
number of baselines 1722 given there should be 1272).
Totally, 2254 sources were observed, more the half of them during the VLBA
Calibrator Survey program.

\begin{figure}
\centering
\hbox{
\epsfxsize=0.42\textwidth \epsfbox{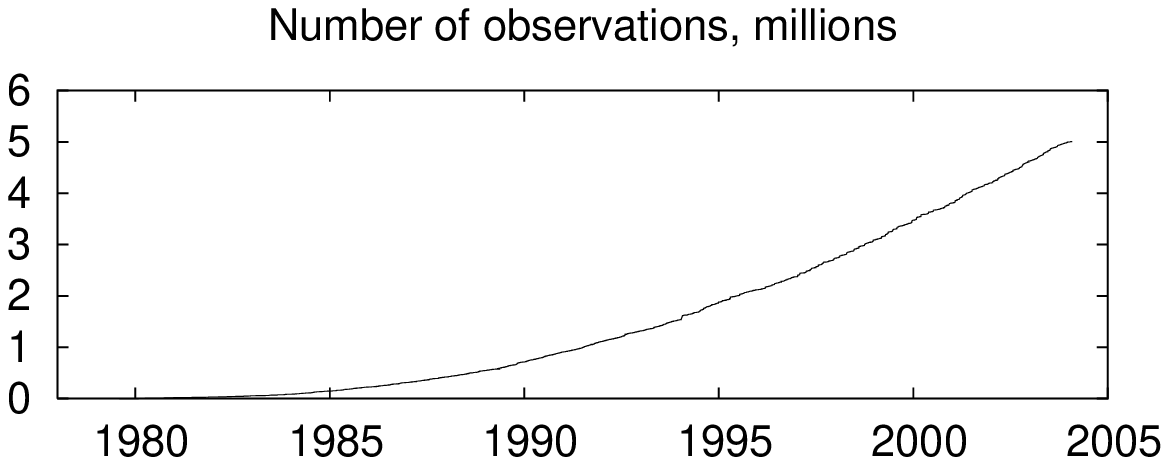}
\epsfxsize=0.42\textwidth \epsfbox{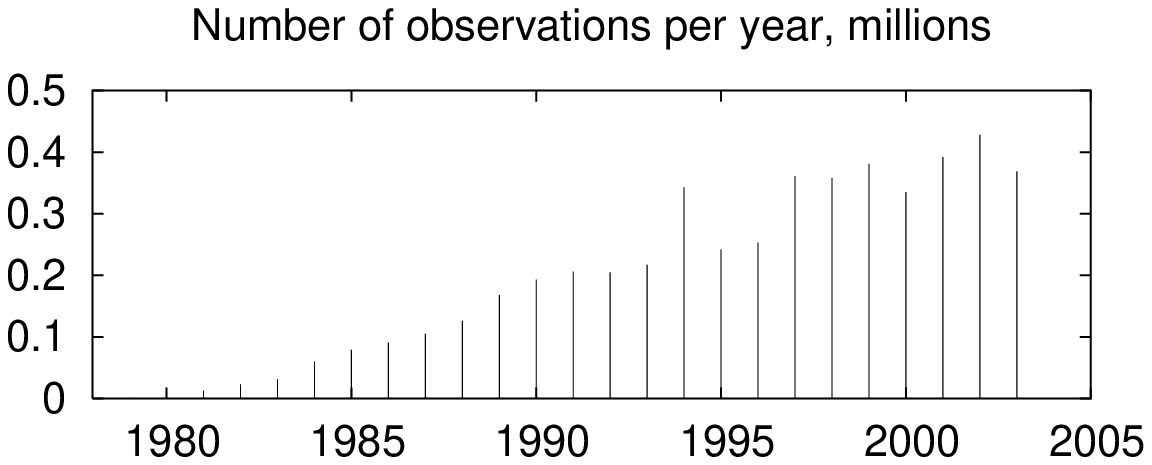}
}
\hbox{
\epsfxsize=0.42\textwidth \epsfbox{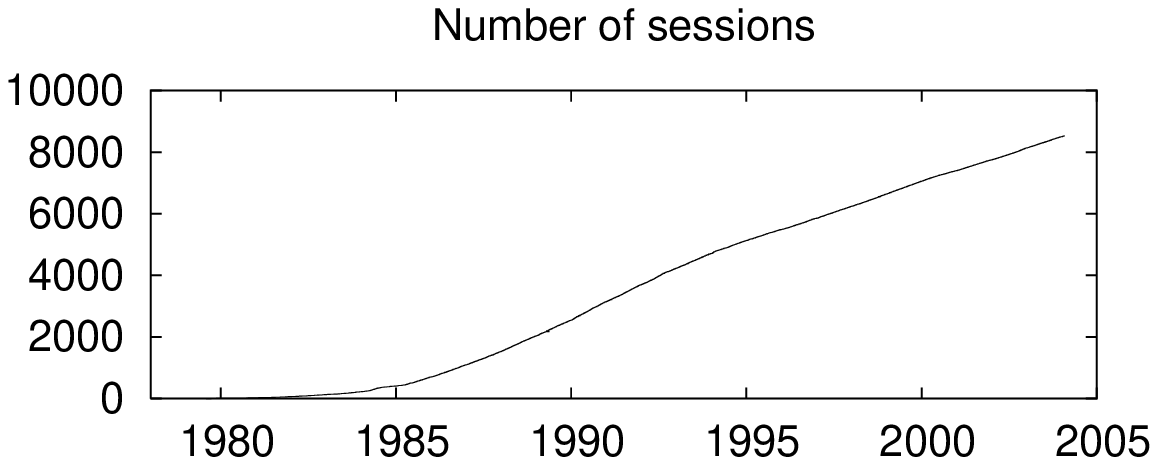}
\epsfxsize=0.42\textwidth \epsfbox{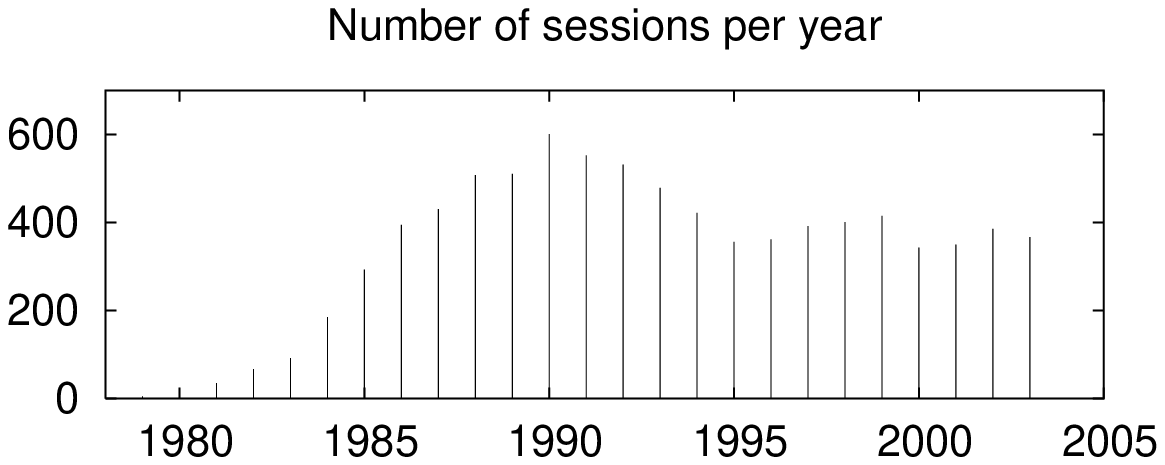}
}
\hbox{
\epsfxsize=0.42\textwidth \epsfbox{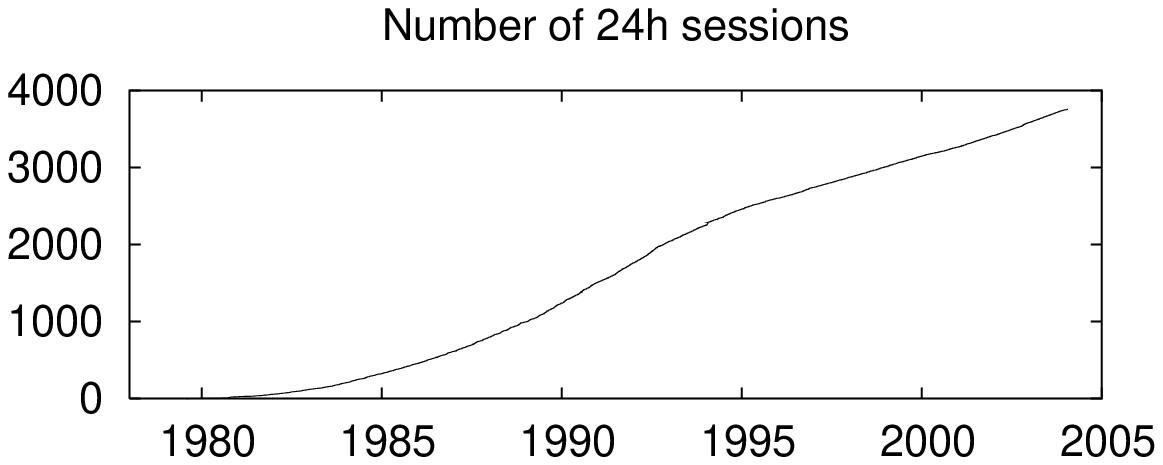}
\epsfxsize=0.42\textwidth \epsfbox{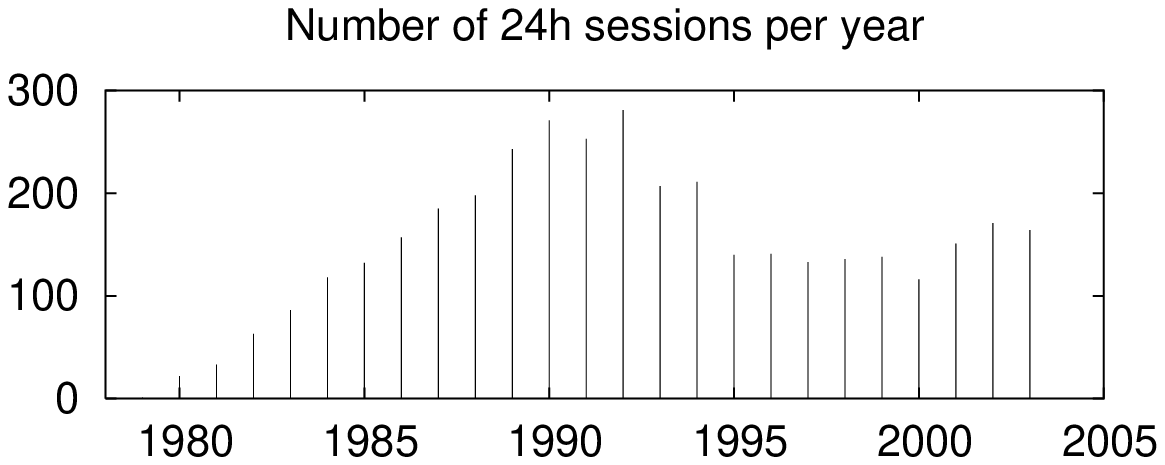}
}
\hbox{
\epsfxsize=0.42\textwidth \epsfbox{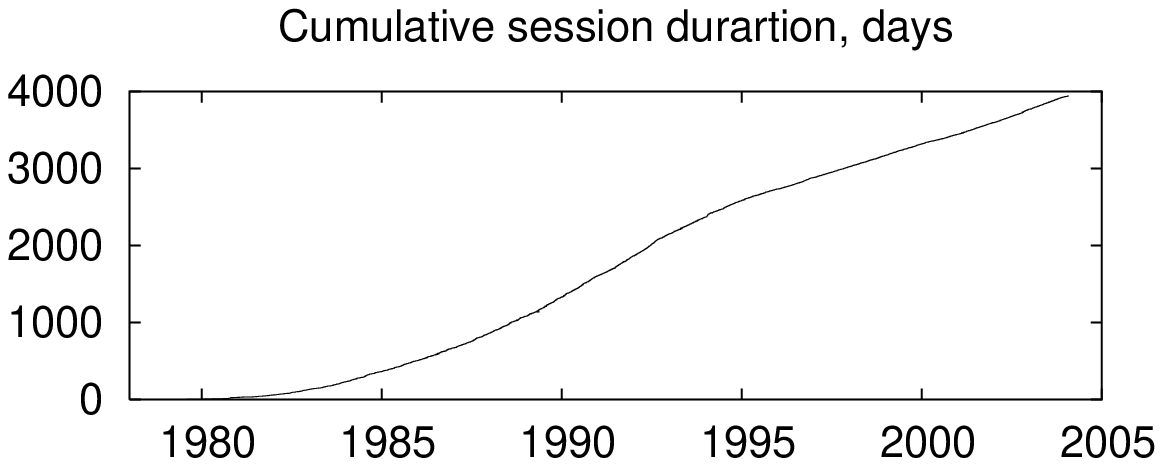}
\epsfxsize=0.42\textwidth \epsfbox{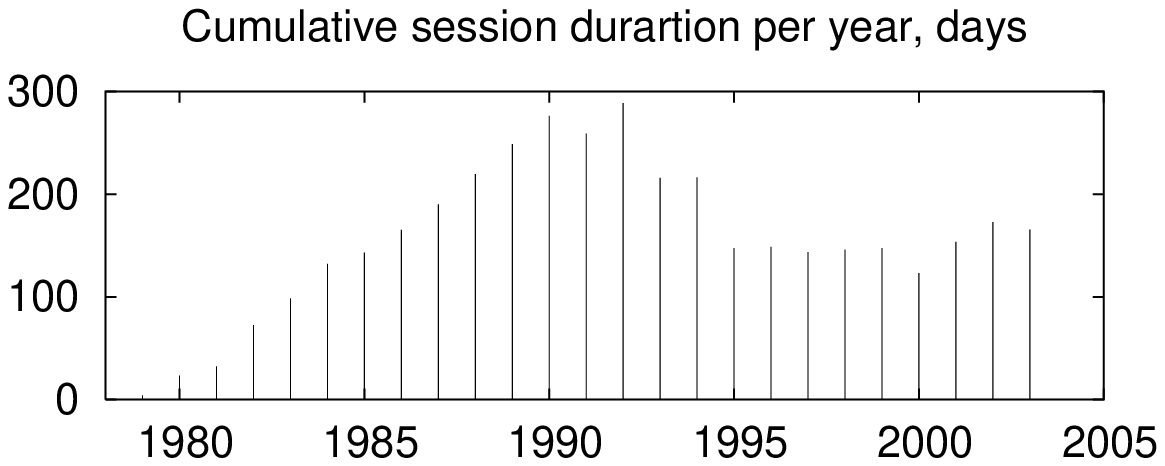}
}
\hbox{
\epsfxsize=0.42\textwidth \epsfbox{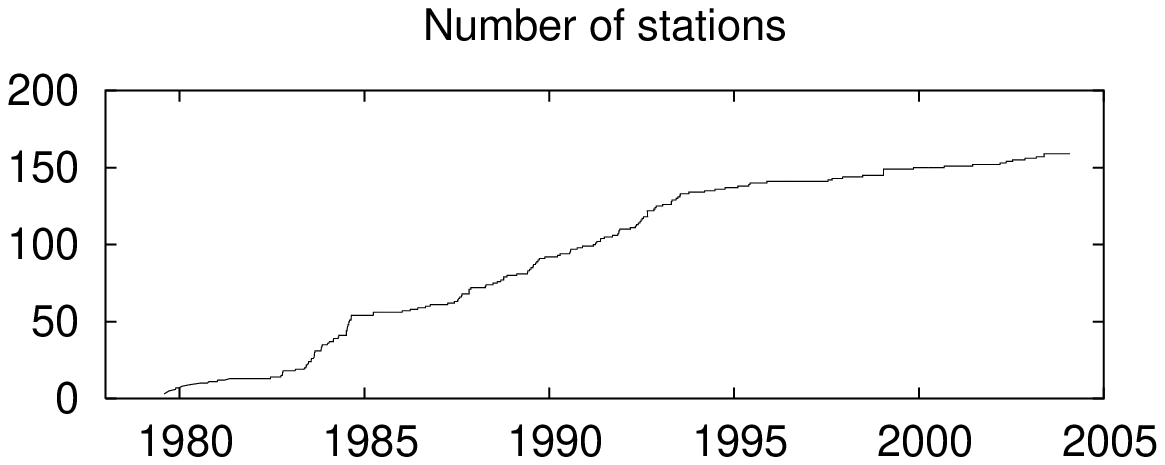}
\epsfxsize=0.42\textwidth \epsfbox{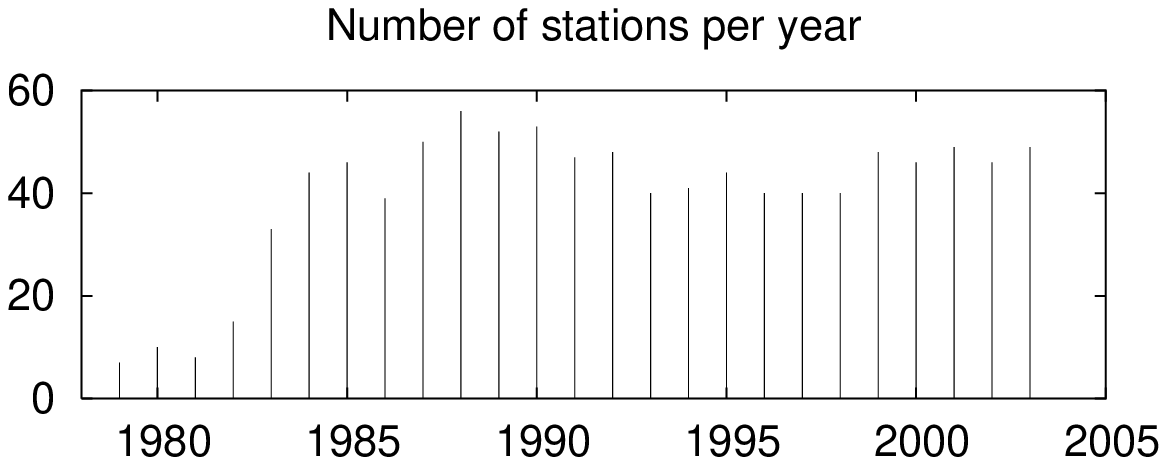}
}
\hbox{
\epsfxsize=0.42\textwidth \epsfbox{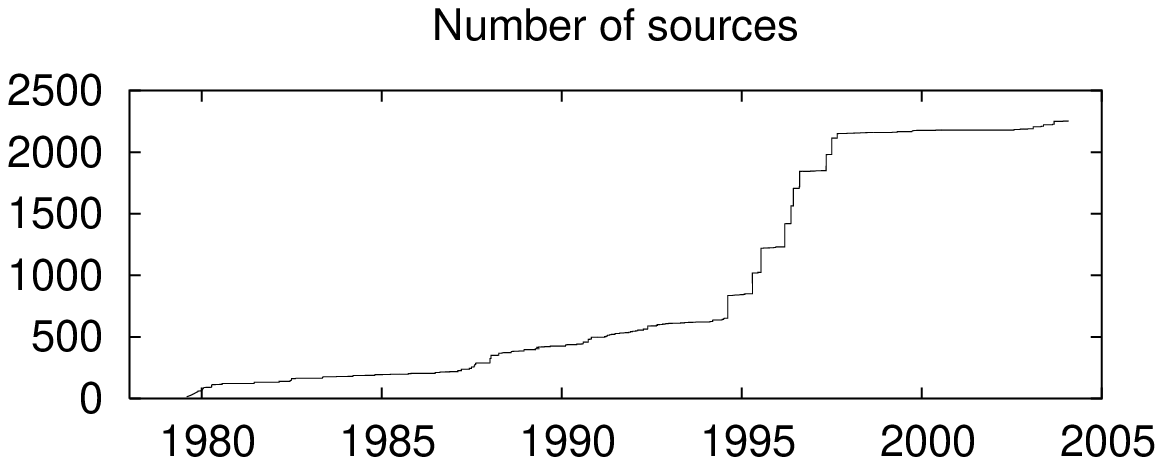}
\epsfxsize=0.42\textwidth \epsfbox{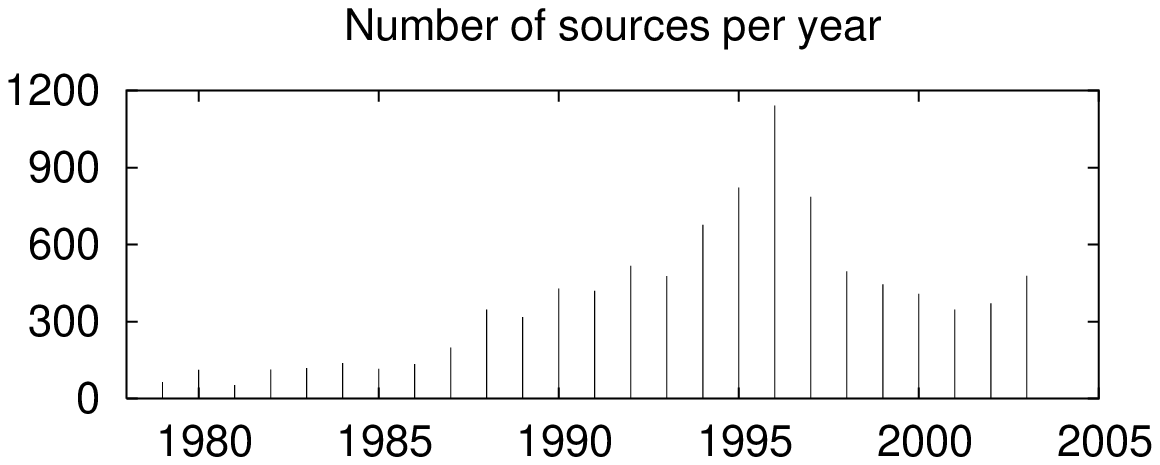}
}
\centerline{\epsfxsize=0.42\textwidth \epsfbox{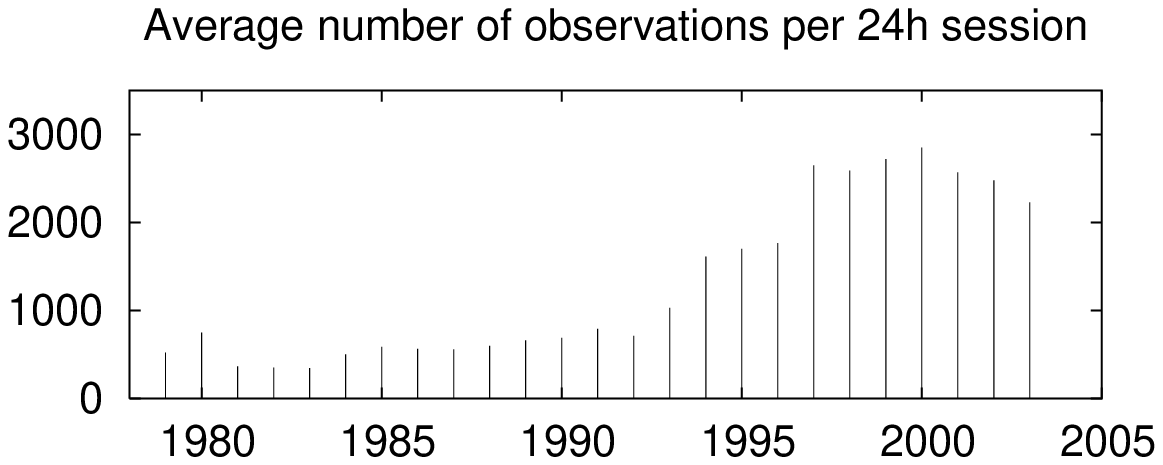}}
\caption{Overall statistics.}
\label{fig:statall}
\end{figure}

Figure~\ref{fig:statall} shows how the overall result was reached.
It is interesting to see how much time was needed to get each million
observations (Table~\ref{tab:millions}).
Apparently, a limit of the capacity
of existing IVS network is reached in the late 1990s.

\begin{table}[ht!]
\centering
\caption{Time taken to collect each million observations ($\Delta T$),
the corresponding period of observations ($T_{beg}\!-\!T_{end}$), and
number of sessions.
Some inconsistencies between $\Delta T$ and $T_{beg}\!-\!T_{end}$
are due to rounding.}
\label{tab:millions}
\small
\vspace{5mm}
\begin{tabular}{|l|c|c|c|c|c|}
\hline
Million           &  1st & 2nd & 3rd & 4th & 5th \\
\hline
$\Delta T$, years & 11.9 & 3.9 & 3.3 & 2.6 & 2.6 \\
$T_{beg}\!-\!T_{end}$ & 1979.6--1991.5 & 1991.5--1995.5 & 1995.5--1998.7 &
                    1998.7--2001.4 & 2001.4--2004.0 \\
\# of sessions     & 3425 & 1873 & 1234 & 1002 & ~976 \\
\# of 24h sessions & 1628 & ~901 & ~444 & ~348 & ~432 \\
\hline
\end{tabular}
\end{table}

\vspace{-5mm}
\noindent
\begin{tabular}{lrl}
&& \\[2ex]
Here are some extreme statistics with examples: && \\[0.5ex]
Maximum number of stations:               &     20  & (991220XA) \\
Maximum number of baselines:              &    188  & (991220XA) \\
Maximum number of sources:                &    263  & (950715XV) \\
Maximum number of observations:           &  34221  & (991220XA) \\
Maximum number of good observations:      &  30372  & (991220XA) \\
Maximum number of bad observations:       &   4092  & (940812XV) \\
Maximum percentage of good observations:  &  100.0  & (860223X~) \\
Maximum percentage of bad observations:   &   90.8  & (911205MV) \\
The longest sessions, h:                  &   99.2  & (830520D~) \\
\end{tabular}
\bigskip

Table~\ref{tab:stations} shows most active stations during the whole period
of observations, and Tables~\ref{tab:sources}--\ref{tab:baselines}
present statistics for sources and baselines.
The longest attempted baselines is SESHAN25--TIGOCONC (12660~km), but no
successful observations (zero quality code in NGS files) was obtained.
The longest baseline with successful observations is DSS65--HOBART26
(12520~km).  The shortest baselines was KAUAI--KOKEE (39~m)

\begin{table}[ht!]
\centering
\caption{Most active stations, Nobs~-- number of observations,
 Nsess~-- number of sessions.}
\label{tab:stations}
\small
\begin{tabular}{|lc|lc|lc|lc|}
\hline
\multicolumn{4}{|c|}{Ordered by number of observations} &
\multicolumn{4}{|c|}{Ordered by number of sessions} \\
\cline{1-8}
\multicolumn{2}{|c|}{All sessions} & \multicolumn{2}{|c|}{24h sessions} &
\multicolumn{2}{|c|}{All sessions} & \multicolumn{2}{|c|}{24h sessions} \\
\cline{1-8}
~~Station& Nobs    & ~~Station& Nobs    & ~~Station& Nsess & ~~Station& Nsess \\
\hline
GILCREEK & 1016731 & GILCREEK & 1007188 & WETTZELL &  6291 & WETTZELL & 1893 \\
WETTZELL & ~979313 & WETTZELL & ~916828 & WESTFORD &  3849 & GILCREEK & 1731 \\
WESTFORD & ~876858 & WESTFORD & ~853711 & GILCREEK &  1937 & WESTFORD & 1640 \\
KOKEE    & ~541291 & KOKEE    & ~529154 & KOKEE    &  1639 & KOKEE    & ~953 \\
NYALES20 & ~410231 & NYALES20 & ~409257 & NRAO20   &  1389 & RICHMOND & ~734 \\
MATERA   & ~314207 & MATERA   & ~311260 & NRAO85 3 &  1145 & HRAS 085 & ~732 \\
ONSALA60 & ~292715 & ONSALA60 & ~291044 & RICHMOND &  ~794 & MOJAVE12 & ~726 \\
MOJAVE12 & ~288671 & MOJAVE12 & ~286644 & HRAS 085 &  ~742 & FORTLEZA & ~704 \\
LA-VLBA  & ~268754 & LA-VLBA  & ~268754 & MOJAVE12 &  ~737 & HARTRAO  & ~593 \\
ALGOPARK & ~260134 & ALGOPARK & ~260134 & FORTLEZA &  ~706 & ALGOPARK & ~527 \\
\hline
\end{tabular}
\end{table}

\begin{table}[ht!]
\centering
\caption{Most observed sources, Nobs~-- number of observations,
 Nsess~-- number of sessions.}
\label{tab:sources}
\small
\begin{tabular}{|lc|lc|lc|lc|}
\hline
\multicolumn{4}{|c|}{Ordered by number of observations} &
\multicolumn{4}{|c|}{Ordered by number of sessions} \\
\cline{1-8}
\multicolumn{2}{|c|}{All sessions} & \multicolumn{2}{|c|}{24h sessions} &
\multicolumn{2}{|c|}{All sessions} & \multicolumn{2}{|c|}{24h sessions} \\
\cline{1-8}
~~Source & Nobs    & ~~Source & Nobs    & ~~Source & Nsess & ~~Source & Nsess \\
\hline
0552+398 &  271721 & 0552+398 &  266837 & 0552+398 &  5996 & 0552+398 & 3517 \\
4C39.25  &  201307 & 4C39.25  &  198602 & 4C39.25  &  5137 & 4C39.25  & 3231 \\
0059+581 &  166093 & 0059+581 &  163363 & 0059+581 &  4299 & OJ287    & 3099 \\
1803+784 &  146816 & 1803+784 &  138602 & 1803+784 &  4201 & 0528+134 & 2904 \\
OJ287    &  135129 & OJ287    &  131791 & OJ287    &  3765 & 1741-038 & 2640 \\
1739+522 &  131596 & 1739+522 &  131196 & 1739+522 &  3333 & 0727-115 & 2566 \\
1357+769 &  125816 & 1357+769 &  122314 & 1357+769 &  3130 & 0229+131 & 2219 \\
0528+134 &  113163 & 0528+134 &  112026 & 0528+134 &  2894 & 1803+784 & 2162 \\
1308+326 &  101217 & 1741-038 &  ~99174 & 1741-038 &  2814 & 1334-127 & 2139 \\
1741-038 &  ~99965 & 1308+326 &  ~99173 & 1308+326 &  2717 & 2145+067 & 2046 \\
\hline
\end{tabular}
\end{table}

\begin{table}[ht!]
\centering
\caption{Most observed baselines, Nobs~-- number of observations,
 Nsess~-- number of sessions.}
\label{tab:baselines}
\small
\begin{tabular}{|lc|lc|}
\hline
\multicolumn{4}{|c|}{Ordered by number of observations} \\
\cline{1-4}
\multicolumn{2}{|c|}{All sessions} & \multicolumn{2}{|c|}{24h sessions} \\
\cline{1-4}
\multicolumn{1}{|c}{Baseline} & Nobs & \multicolumn{1}{|c}{Baseline} & Nobs \\
\hline
WESTFORD  WETTZELL &  171093 & WESTFORD  WETTZELL &  149602 \\
GILCREEK  WESTFORD &  134234 & GILCREEK  WESTFORD &  133942 \\
GILCREEK  KOKEE    &  116313 & GILCREEK  KOKEE    &  116290 \\
GILCREEK  WETTZELL &  111450 & GILCREEK  WETTZELL &  111145 \\
NYALES20  WETTZELL &  ~90820 & NYALES20  WETTZELL &  ~90703 \\
HRAS 085  WESTFORD &  ~70704 & HRAS 085  WESTFORD &  ~70233 \\
KOKEE     WETTZELL &  ~65154 & MOJAVE12  WESTFORD &  ~63085 \\
MOJAVE12  WESTFORD &  ~63252 & FORTLEZA  WETTZELL &  ~62451 \\
FORTLEZA  WETTZELL &  ~62456 & MATERA    WETTZELL &  ~59681 \\
MATERA    WETTZELL &  ~59945 & GILCREEK  KAUAI    &  ~56883 \\
\hline
\hline
\multicolumn{4}{|c|}{Ordered by number of sessions} \\
\cline{1-4}
\multicolumn{2}{|c|}{All sessions} & \multicolumn{2}{|c|}{24h sessions} \\
\cline{1-4}
\multicolumn{1}{|c}{Baseline} & Nsess & \multicolumn{1}{|c}{Baseline} & Nsess \\
\hline
WESTFORD  WETTZELL &  3312 & WESTFORD  WETTZELL & 1112 \\
KOKEE     WETTZELL &  1415 & GILCREEK  WETTZELL & ~799 \\
NRAO20    WETTZELL &  1297 & KOKEE     WETTZELL & ~732 \\
GILCREEK  WETTZELL &  ~818 & GILCREEK  WESTFORD & ~680 \\
GILCREEK  WESTFORD &  ~683 & HRAS 085  WESTFORD & ~595 \\
NRAO85 3  WETTZELL &  ~647 & GILCREEK  KOKEE    & ~591 \\
HRAS 085  WESTFORD &  ~602 & FORTLEZA  WETTZELL & ~571 \\
GILCREEK  KOKEE    &  ~593 & RICHMOND  WESTFORD & ~566 \\
FORTLEZA  WETTZELL &  ~572 & FORTLEZA  KOKEE    & ~548 \\
RICHMOND  WESTFORD &  ~567 & RICHMOND  WETTZELL & ~538 \\
\hline
\end{tabular}
\end{table}

\begin{figure}[ht!]
\centering
\hbox{
\epsfxsize=0.42\textwidth \epsfbox{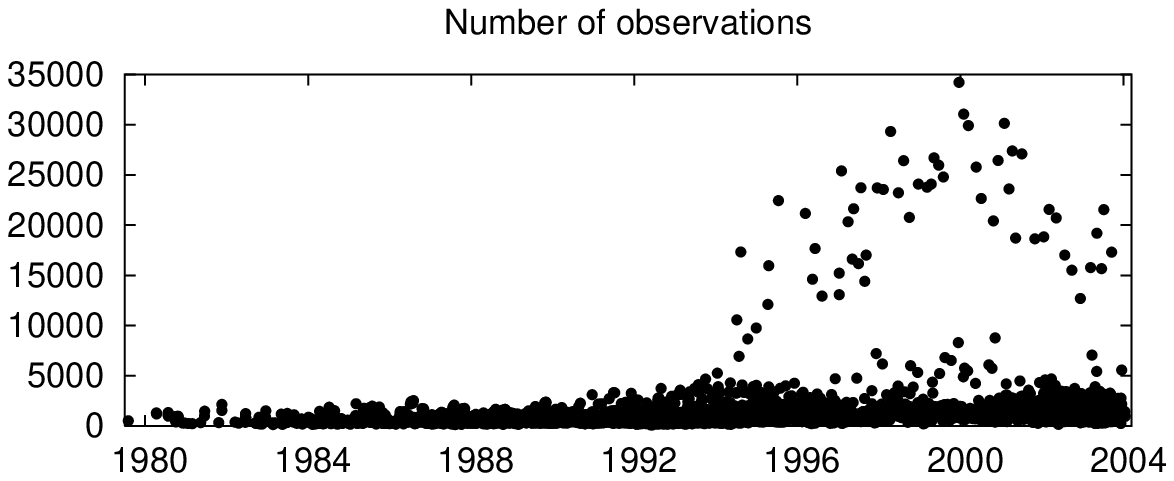}
\epsfxsize=0.42\textwidth \epsfbox{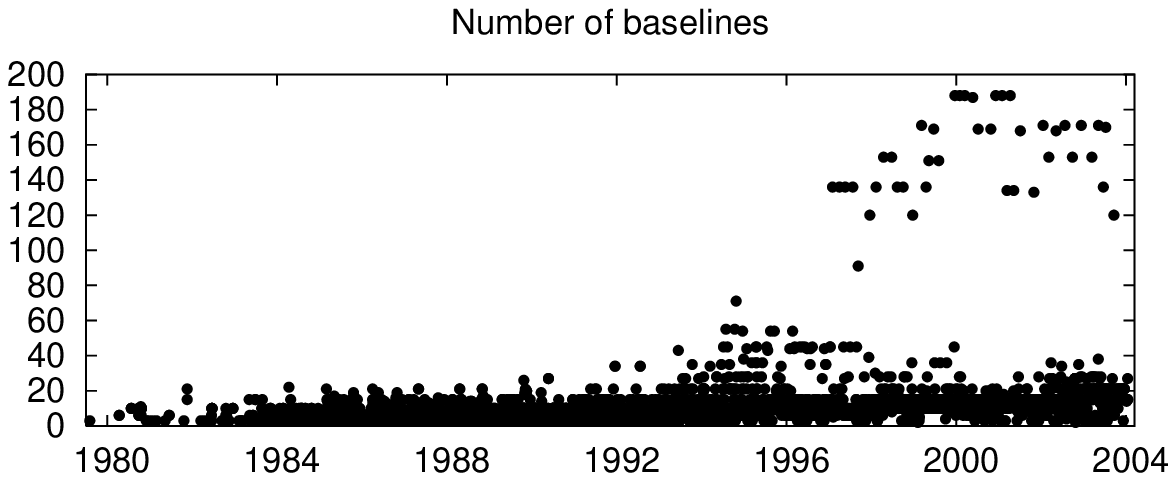}
}
\hbox{
\epsfxsize=0.42\textwidth \epsfbox{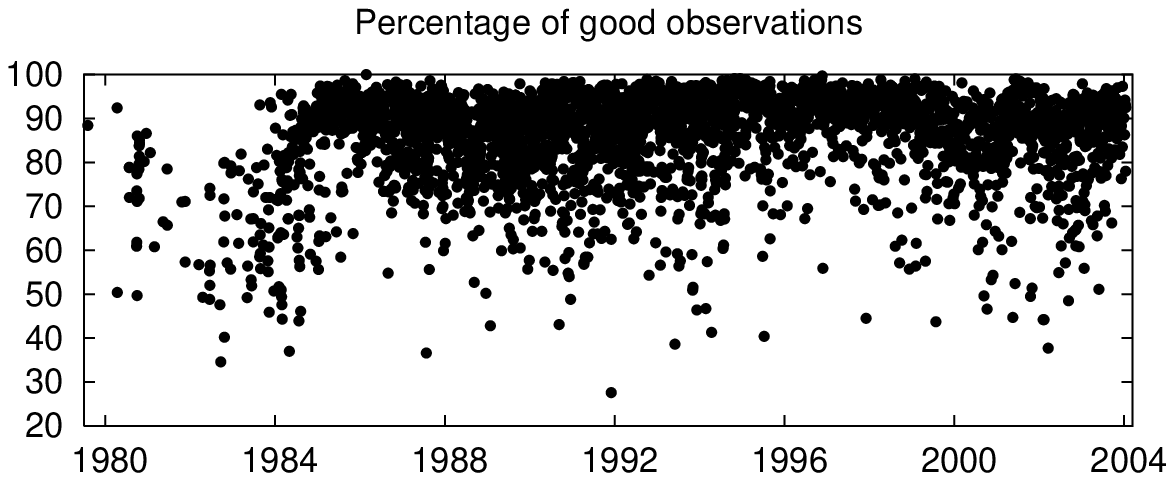}
\epsfxsize=0.42\textwidth \epsfbox{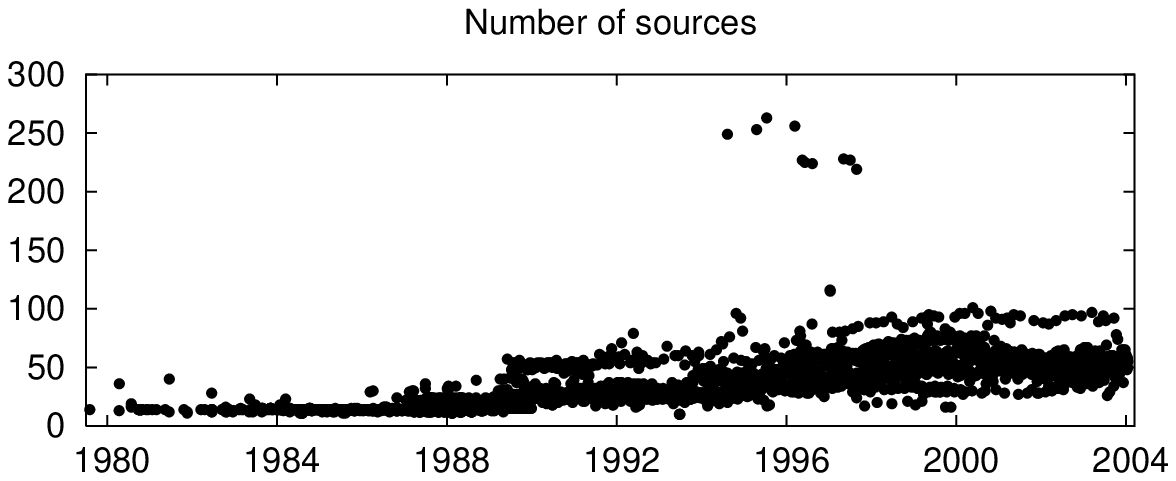}
}
\hbox{
\epsfxsize=0.42\textwidth \epsfbox{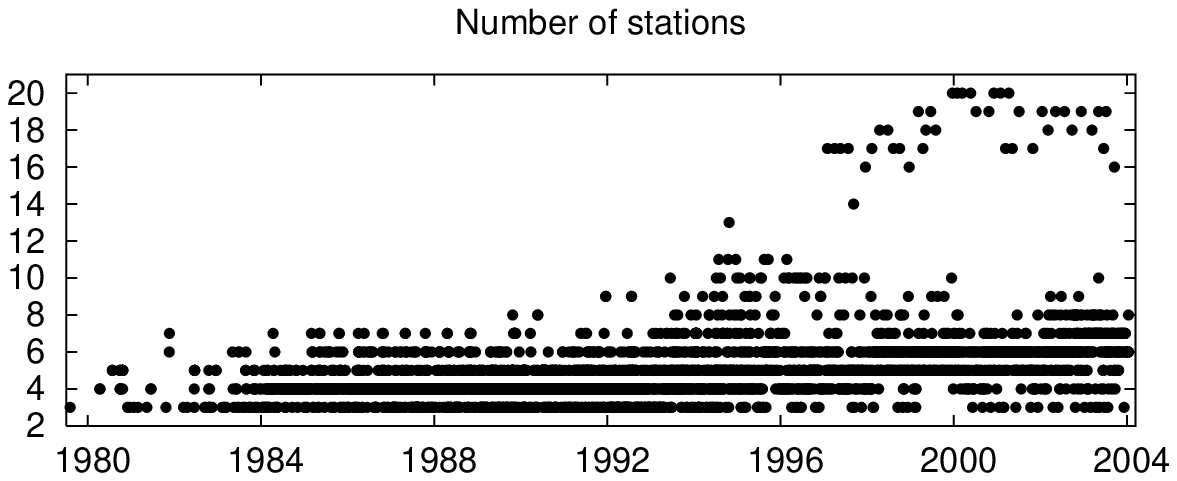}
\epsfxsize=0.42\textwidth \epsfbox{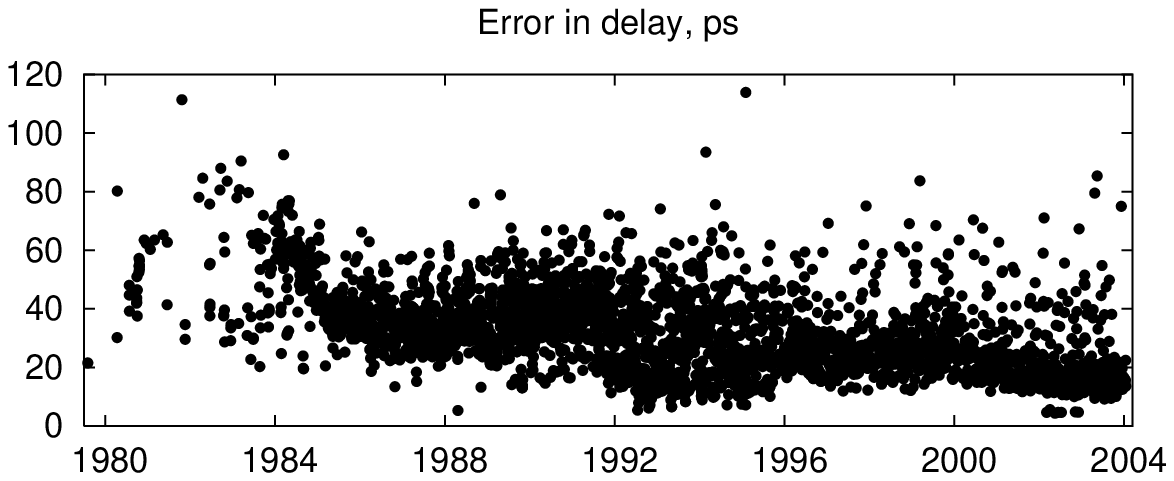}
}
\caption{Statistics of the observational data. Each point corresponds to one
24h session. Sessions with the largest numbers of stations/baselines were
observed along with VLBA network. Sessions with the largest number of sources
were observed in the framework of the VLBA Calibration Survey program.}
\label{fig:statdata}
\end{figure}

\begin{figure}[ht!]
\centering
\hbox{
\epsfxsize=0.42\textwidth \epsfbox{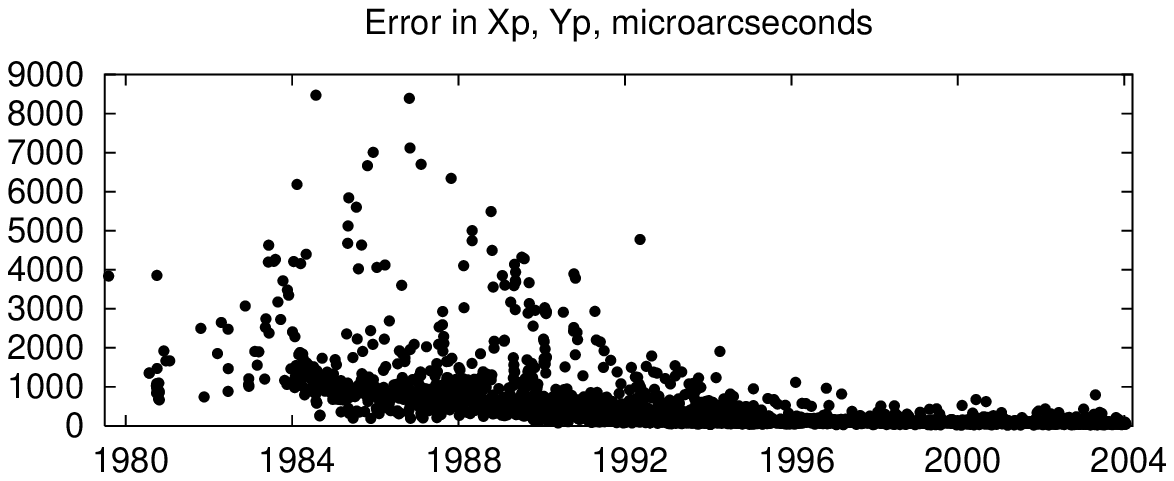}
\epsfxsize=0.42\textwidth \epsfbox{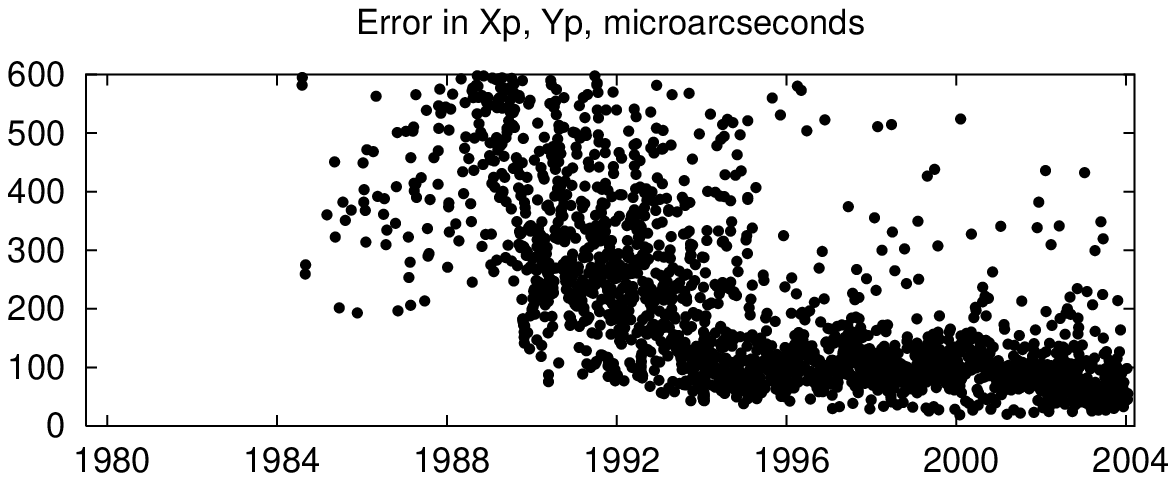}
}
\hbox{
\epsfxsize=0.42\textwidth \epsfbox{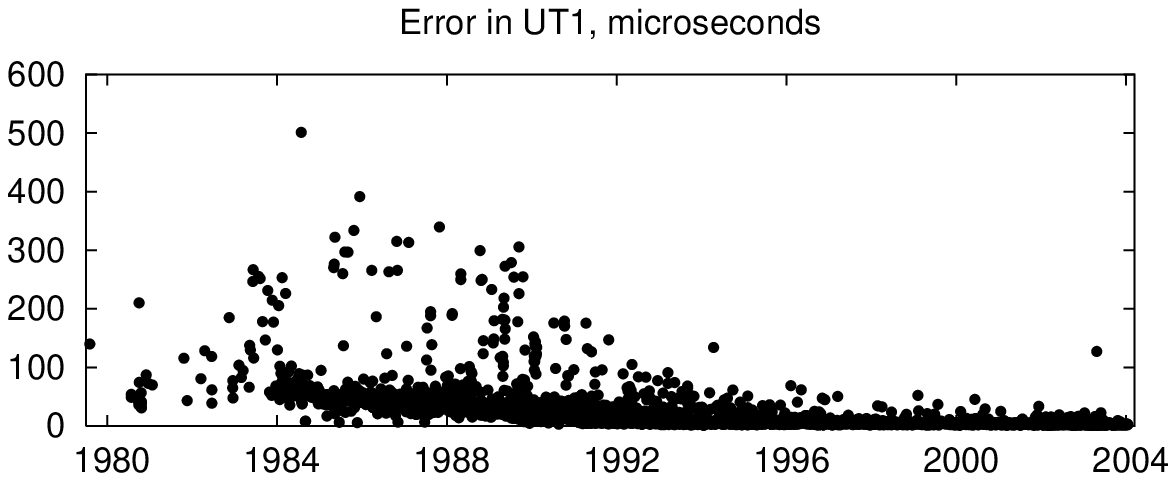}
\epsfxsize=0.42\textwidth \epsfbox{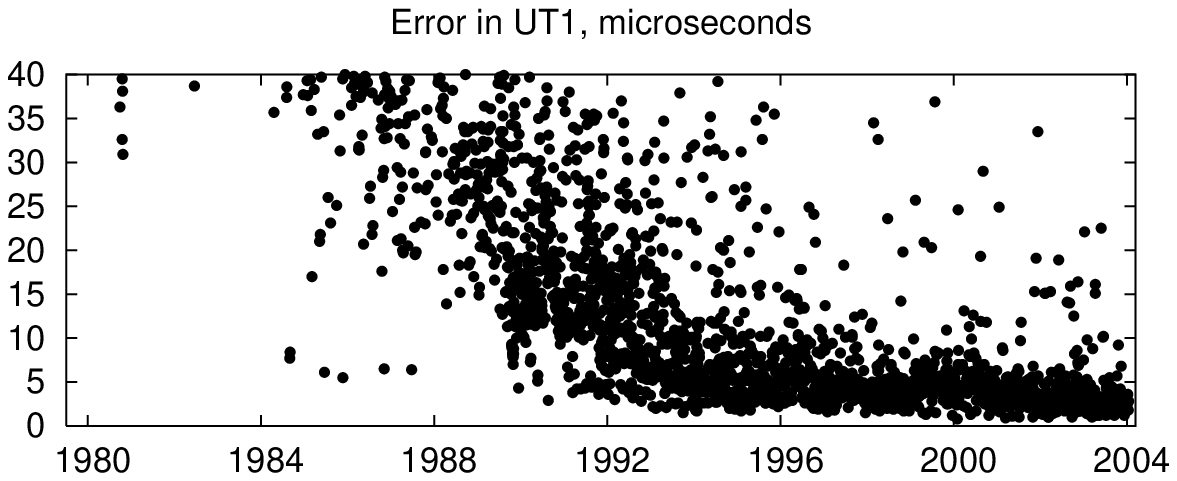}
}
\hbox{
\epsfxsize=0.42\textwidth \epsfbox{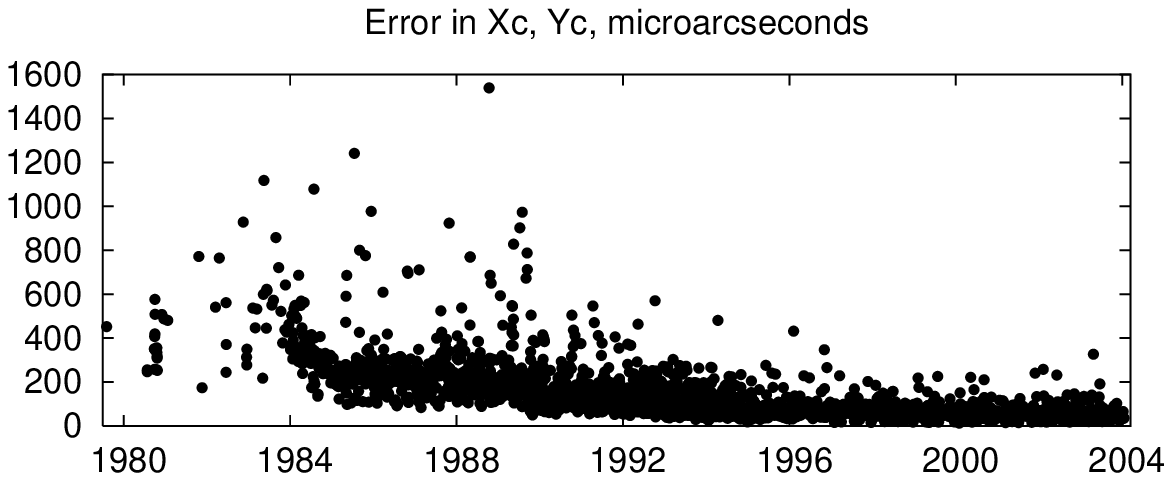}
\epsfxsize=0.42\textwidth \epsfbox{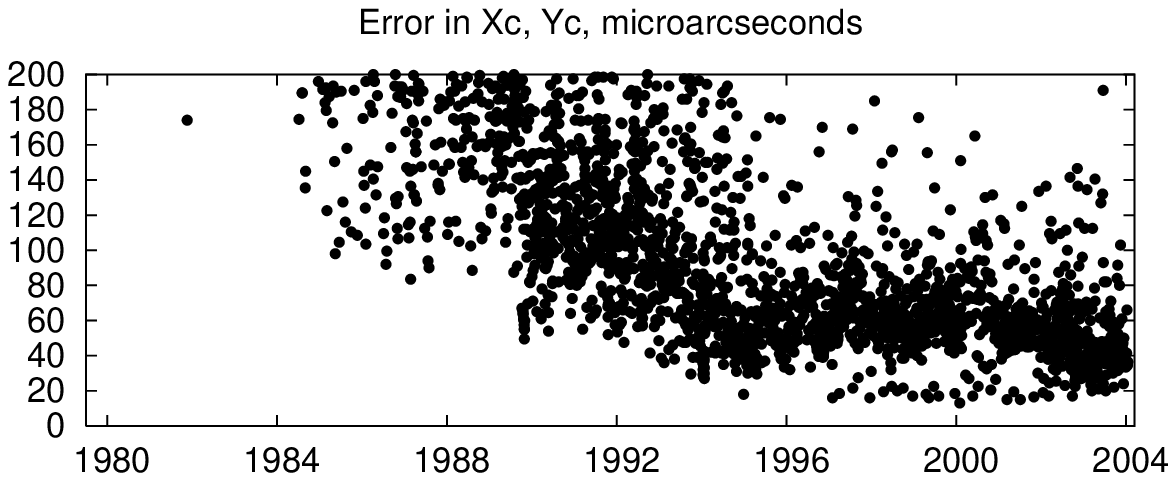}
}
\caption{Errors in EOP (on the left are zoomed data). Each point corresponds
to one 24h session. Note clear correlation with improvement in delay
precision, and also with number of stations/baselines
(Figure~\ref{fig:statdata}).}
\label{fig:eoperr}
\end{figure}

Figures~\ref{fig:statdata} and \ref{fig:eoperr} show the evolution
of some observational data and EOP uncertainty (IAA EOP series) with time.

\end{document}